# A Parallel Genetic Algorithm for Generalized Vertex Cover Problem

Drona Pratap Chandu
*Amazon Development Centre India,*
*Chennai, India*

*Abstract*— This paper presents a parallel genetic algorithm for generalised vertex cover problem( GVCP) using Hadoop Map-Reduce framework. The proposed Map-Reduce implementation helps to run the genetic algorithm for generalized vertex cover problem(GVCP) on multiple machines parallely and computes the solution in relatively short time.

## I. INTRODUCTION

The vertex cover problem (VCP) for undirected graph $G=(V,E)$ is to find find a subset $S$ of $V$ such that for each edge $e$ in E at least one of end points of $e$ belongs to $S$ and $|S|$ is as small as possible. VCP was proved as a NP-complete problem by Karp[5] in 1972. There are many variants of generalized vertex cover problem. In this paper generalized vertex cover problem formulated by Hassin and Levin[1] is considered. Marija[2] proposed a genetic algorithm for the generalized vertex cover problem proposed by Hassin and Levin[1].

This paper presents a parallel implementation of genetic algorithm for generalized vertex cover problem (GVCP) using Hadoop Map-Reduce framework. The proposed Map-Reduce implementation helps to run the genetic algorithm for generalized vertex cover problem on multiple machines parallely and computes the solution in a relatively short time.

## II. OBJECTIVE FUNCTION

Let $G = (V,E)$ be an undirected graph. For every edge $e \in E$ three numbers $d_0(e) \geq d_1(e) \geq d_2(e) \geq 0$ are given and for every vertex $v \in V$ a number $c(v) \geq 0$ is given.

For a subset $S \subseteq V$ denote $S_1 = V \setminus S$, $E(S)$ is the set of edges whose both end-vertices are in $S$, $E(S,S_1)$ is the set of edges that connect a vertex from $S$ with a vertex from $S_1$, $c(S) = \Sigma c(v)$, and for i = 0, 1, 2 $d_i(S) = \Sigma d_i(e)$ and $d_i(S, S_1) = \Sigma d_i(e)$ $e \in E(S,S_1)$. The generalized vertex cover problem is to find a vertex set $S \subseteq V$ that minimizes the *cost* $c(S) + d_2(S) + d_1(S, S_1) + d_0(S_1)$.

**Example 1.** Let G=(V,E) with V={1,2,3,4} and E= {(1,2),(1,3),(1,4),(2,3),(3,4)} be the given graph, costs associated with vertices are {10,20,30,40} and costs associated with edges (d0,d1,d2) are given by {(50,30,20)(40,4030)(50,20,20)(30,20,10)(20,20,20)}.

The optimal objective value in this example is 150 and the optimal vertex cover {1} consists of only one vertex. The corresponding vertex cost c(S) = c(1) = 10 and the edge costs are d1(S, S1) = d1(1, 2) + d1(1, 3) + d1(1, 4) = 30 + 40 + 20 = 90,d0(S1) = d0(2, 3) + d0(3, 4) = 30 + 20 = 50 and d2(S) = 0.

## III. HADOOP MAP-REDUCE

Map-Reduce is a programming model developed by Dean and Ghemawat[3] for writing highly scalable parallel programs. User of Map-Reduce framework have to implement two functions map and reduce. The programs written in the Map-Reduce style is automatically parallelized by Map-Reduce run time system. Scheduling, distributing, aggregation of the results and failure recovery was automatically handled by Map-Reduce runtime system. Apache is providing an open source implementation of Map-Reduce as a part of its Hadoop project[4].

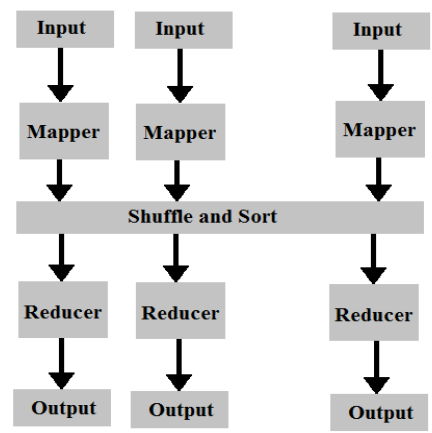

Figure 1 . Hadoop Map-Reduce

The map method takes a key value pair as input and emits key value pairs as output. The output key value pairs of map functions running on multiple mapper machines is collected by runtime system and it sort them and group them according to key value. Then it passes a key and list of values corresponding to that key to reduce function. The reduce function process the given (key ,values list) and outputs (key ,value) pair. Map operations and reduce operations are distributed among several machines in Hadoop Cluster as shown in Figure 1. Many real world problems can be solved using Map-Reduce model.

## IV. A GENETIC ALGORITHM FOR GENERALISED VERTEX COVER PROBLEM (GVCP)

Holland[6] proposed genetic algorithms(GAs) and have been applied to many optimization problems and are proven to be efficient in finding globally optimal solutions. GAs are computational models inspired by evolutionary theory. Genetic algorithms(GA) begins with a random sample of chromosomes where each chromosome represents a solution to the given problem. GA repeatedly applies operators like selection,mutation,crossover that resemble a process of natural evolution to create next generation from selected population of current generation.

GA proposed by Marija[2] can be summarized as below
1. Create *Z* random individuals including four special individuals.
2. Compute fitness of individuals in current generation.
3. Select *E* individuals with high fitness and add them to next generation.
4. Select *Z-E* pairs of individuals and apply cross over , mutation operations on selected individuals to produce a pair of child individuals in the next generation.
5. Repeat steps 2 to 4 for desired number of times.
6. Returns best solution in the last generation

The elements of GA proposed by Marija[2] are described below.

*A. Encoding*

Let *(V1,V2,... Vn)* be the labels of vertices in the given graph. Each solution to GVCP is represented by a binary code of length |V|. Digit 1 in *Kth* position of a binary code indicates the vertex *Vk* is in the solution. Digit 0 in the *Kth* positon of binary code indicates the vertex *Vk* is not in the solution. The solution represented by binary code *B* is set of vertices *Vi* such that *ith* bit in *B* is 1.

**Example 2.** Let the binary code *BC* is 0001 for a graph with vertices *{V1,V2,V3,V4}*, the solution *S* represented by *BC* is *{V4}*.

*B. Population Initialisation*

Initial population is randomly created such that diversity is maximal. Every individual is a random binary string of length $|V|$.

*C. Selection*

Best *E* individuals having higher fitness are selected as elitism which directly proceed into the next generation. To generate remaining population *(Z - E)* parents are selected into the mating pool using fine grained tournament selection described in [7] with desired average tournament size *tSize*. Tournament with size *floor(tSize)* is held *K1* times and tournament with size *ceil(tSize)* is held *K2* times such that overall average tournament size is closer to *tSize*. In each round, it selects *K1 or K2* different individuals from population. One of the selected individuals is added to the mating pool according to probability distribution based on fitness.

*D. Crossover*

To create next generation individuals, crossover and mutation operation is applied on randomly selected two individuals from mating pool. The standard one point crossover is used . Crossover is performed by exchanging segments of two parents genetic codes from a randomly chosen cut point and cross over is performed with probability

$P_{CROSS} = 0.85$.

*E. Mutation*

Mutation changes a randomly selected gene in the individual with certain mutation rate. When all the individuals have same gene on certain position , the gene can't changed by crossover and selection operators. The gene is called frozen. Mutation rate of froze gene is (1.0/N) and of non-frozen ones is (0.4/N).

*F. Local search around best individual*

The proposed genetic algorithm use local search around best individual in the current generation to improve the best individual. The local search is performed by using add/remove heuristic with first improvement, that usually outperforms best improvement heuristic.

In more detail, for each ch ∈ {1, 2,. . . , |V |}, local search tries to complement S[ch] to 1−S[ch]. If that brings improvement to the objective value, the change is performed. That process is repeated until there is no improvement in some iteration, and the last objective value is declared as final objective value.

V. IMPLEMENTATION USING MAP-REDUCE

The proposed model requires to run the Map-Reduce iteratively. Each iteration takes population of one generation as input and applies necessary evolutionary operators during Map-Reduce to produce the next generation as output. Only one mapper is used and number of reducers can be increased to compute the solution quickly. Overall process can be summarised as follows
1. Create initial population.
2. Compute fitness of each chromosome in the initial population.
3. Run Map-Reduce with (solution,fitness) as input.
4. Collect output (solution,fitness) pairs of reduce phase.
5. Repeat steps 3 to 4 using output collected in step 4.
6. Output the solution with best fitness in last iteration.

Map function receives (solution,fitness) pairs as input. During Map phase precomputed fitnesses of solutions are used to select the solutions by tournament selection. Solutions and their fitnesses are stored in buffer until all chromosomes are received by Map. Solutions are sorted by fitness and E solutions having best fitness are added to next generation without applying evolutionary operators. For these elite E chromosomes map function outputs a pair of chromosomes first one is current chromosome in the elite E and second is empty string representing illegal chromosome.

To generate the rest of the population of next generation, pairs of chromosomes are selected and emitted as output to be used by the reduce phase. The pairs of chromosomes generated by map function are used as parent chromosomes by reduce function to produce next generation chromosomes.

During the reduce phase evolutionary operators are applied to generate child chromosomes from pair of chromosomes emitted by map phase. When reduce function receives a pair of chromosomes with second chromosome is illegal, reduce just emits the first chromosome and its fitness as output. Best Match Heuristic Packaging Procedure presented in section V is used to compute the fitness of solutions. When reduce function receives a pair of two legal chromosomes, it applies crossover operator to produce two child chromosomes. Then mutation operator is applied over two child chromosomes. The reduce emits (solution,fitness) pairs as output .

In the proposed implementation selection of chromosomes to be added to mating pool runs on single machine as it is configured to have only machine in the map phase. But the process of selection is just picking random numbers and emitting corresponding chromosomes as output and it's execution time is small.

Fitness computation operations,crossover operations, and mutation operations are distributed among all the machines in in reduce phase. This parallel implementation can compute vertex cover in a relatively short time as it runs parallely on multiple machines and running time decreases as the number of machines in reduce phase are increased.

Implementation of Mapper using Java programming language is shown in Figure 2.

```
package ParallelGVCP;
import java.io.IOException;
import java.util.Arrays;
import org.apache.hadoop.io.*;
import org.apache.hadoop.mapred.*;

public class GVCPMapper extends MapReduceBase implements
        Mapper<Text, DoubleWritable,Text, Text>
{  int populationSize = 150;
   int processedItems = 0;
   double[] fitnessVector = new double[populationSize];
   String  population[] = new String[populationSize];
   Text parentOne = new Text("");
   Text parentTwo = new Text("");
   private int E = 50;

   @Override
   public void map(Text key, DoubleWritable value,OutputCollector<Text,
           Text> outputCollector, Reporter reporter) throws IOException
   {   population[processedItems] = key.toString();
      fitnessVector[processedItems]=  value.get();
      processedItems++;
      if( processedItems == populationSize)
      {   MapHelper.sortSolDescendingByFitness (population,fitnessVector);
         population[0] = MapHelper.localSearch(population[0]);
         parentTwo.set("");

         for(int i=0;i<E;i++)
         {  parentOne.set(population[i]);
            outputCollector.collect(parentOne, parentTwo);
         }
         for(int i=0;i < populationSize - E;i++)
         {   String par1 = MapHelper.selectRandom (population,fitnessVector);
             String par2 = MapHelper.selectRandom (population,fitnessVector);
            parentOne.set(par1);
            parentTwo.set(par2);
            outputCollector.collect(parentOne, parentTwo);
         }
         processedItems = 0;
         Arrays.fill(population, null);
      }
   }
}
```

**Figure 2. Implementation of Mapper using Java**

Implementation of Reducer using Java programming language is shown in Figure 3.

```
package ParallelGVCP;
import java.io.IOException;
import java.util.Iterator;
import org.apache.hadoop.io.*;
import org.apache.hadoop.mapred.*;

public class GVCPReducer extends MapReduceBase implements
        Reducer<Text, Text, Text, DoubleWritable>
{ private Text outputKey = new Text();
  private DoubleWritable outputValue = new DoubleWritable();

  @Override
  public void reduce(Text solution, Iterator<Text> solution2,
  OutputCollector<Text, DoubleWritable> results, Reporter reporter)
          throws IOException
  {   String parent1 = solution.toString();
    String parent2 = solution2.next().toString();
    String child1 = null;
    String child2 = null;
    if(parent2.length() == 0)
    {    child1 = parent1;
    } else
    {    int cutpoint = (int) (Math.random() * parent1.length());
      child1 = ReduceHelper.firstChildByCrossover
             (parent1,parent2,cutpoint);
      child2 = ReduceHelper.secondChildByCrossover(parent1,
             parent2,cutpoint);
      child1 = ReduceHelper.mutate(child1);
      child2 = ReduceHelper.mutate (child2);

    }
    double fitness = ReduceHelper.computeFitness (child1);
    outputKey.set(child1);
    outputValue.set(fitness);
    results.collect(outputKey, outputValue);

    if(child2 != null){
       fitness = ReduceHelper.computeFitness (child2);
       outputKey.set(child2);
       outputValue.set(fitness);
       results.collect(outputKey, outputValue);
    }
  }
}
```

**Figure 3. Implementation of Reducer using Java**

The Map-Reduce execution is repeated until convergence condition is satisfied. Output of reduce phase in current iteration is used as input to map phase of next iteration. In every iteration one new generation is created. The best computed solution is represented by best individual in the last generation.

## VI. CONCLUSIONS

This paper presents a parallel genetic algorithm for generalized vertex cover problem using Hadoop Map-Reduce framework. In this implementation fitness computation operations,crossover operations, and mutation operations are distributed among all the machines in Hadoop cluster running reduce phase. This parallel implementation provides a method to compute vertex cover in a relatively short time by running parallely on multiple machines and running time decreases as the number of machines in Hadoop cluster running reduce phase are increased.

## ACKNOWLEDGEMENT

I would like to thank all my colleagues at Amazon India Development Centre, my colleagues at Freescale Semiconductor for their participation in long discussions with me and all my classmates at Indian Institute of Technology Roorkee who used to participate in discussions with me.

## AUTHOR

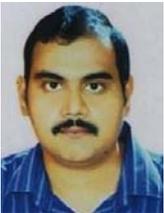

**Drona Pratap Chandu** received Master of Technology in Computer Science and Engineering from Indian Institute of Technology Roorkee, India. He worked as software engineer at Freescale semiconductor India and also he worked as software development engineer at Amazon development centre India recently .